\documentclass[a4paper]{jpconf}
\usepackage{graphicx}
\begin{document}
\title{Schwarzschild-de Sitter black hole in canonical
quantization}
\author{Hossein Ghaffarnejad}
\address{Faculty of Physics, Semnan University, Semnan, IRAN, 35131-19111}
\ead{ hghafarnejad@semnan.ac.ir}
\begin{abstract}
We solve Wheeler-De Witt (WDW) metric probability wave equation on
the apparent horizon hypersurface of the Schwarzschild de Sitter
(SdS) black hole. To do so we choose radial dependent mass
function $M(r)$ for its internal regions in the presence of a
dynamical massless quantum matter scalar field $\psi(r)$ and
calculate canonical supper hamiltonian constraint on $t-$constant
hypersurface near the horizon $r=M(r)$. In this case $M(r)$ become
geometrical degrees of freedom while $\psi(r)$ is matter degrees
of freedom of the apparent horizon. However our solution is
obtained versus the quantum harmonic oscillator which defined
against the well known hermit polynomials. In the latter case we
obtain quantized mass of the SdS quantum black hole as
$\sqrt{\Lambda}M(n)=\big(\frac{2n+1}{12\sqrt{2}}\big)^{\frac{1}{3}}$
in which $\Lambda$ is the cosmological constant and
$n=0,1,2,\cdots $ are quantum numbers of the hermit polynomials.
This shows that a quantized SdS in its ground state has a nonzero
value for the mass $M(0)=0.38914/\sqrt{\Lambda}$. Thus one can
infer that the latter result satisfies Penrose hypotheses for
cosmic censorship where a causal singularity may be covered by a
horizon surface.
\end{abstract}
\section{Introduction}
 Quantum instability of the black holes was discovered firstly by Hawking [1,2]. Then some authors
 studied more details of the quantum black holes evaporation. Properly a
detailed picture of the evaporation process and its final state
can be given only within the framework of a complete and
self-consistent pure quantum gravity theory, which has not yet to
be found [3]. It will be valid for the Planck scale of the
universe [4,5] \footnote{$M_p=(\hbar c/G)^{1/2}=2.18\times10^{-8}
kg;\newline~~~~~D_p=(\hbar G/c^3)^{1/2}=1.62\times 10^{-35}
m,\newline~~~~~T_p=\hbar/M_p c^2=5.31\times10^{-44} s $}. In
absence of a pure quantum gravity theory, the evaporation process
of quantum black holes and their final structure were studied in
different approaches named as the semiclassical perturbative
[1,2,6,7,8] and the non-perturbative Dirac`s canonical
quantization [9-22] methods respectively.
 At present a tractable way to non-perturbative
quantum gravity is the WDW functional approach of canonical
quantization [17].
 Recent developments in quantum cosmology are based on the
 mini-super-space analysis of the WDW equation. To construct black
 hole mini-super-space models, we can assume spherically symmetric
 metrics. A full canonical formalism of spherically symmetric
 systems was extensively studied by Hajicek et al [14,15]. The
 difficult problem in quantization is how to treat the
 super-momentum constraint condition. Unlike in cosmological
 situations, we cannot find adequate mini-super-space variables
 for which the super-momentum constraint become trivial. For example,
 Rodrigues et al [22], proposed a black hole mini-super-space model to discuss
 the wave function of a decaying black hole. However the WDW wave functional equation
  derived from their mini-super-space variables turned out to be incompatible with the super-momentum
 constraint [23]. Their model is limited to a purely gravitational
 system. Then the dynamical behavior of a spherically symmetric
 metric is severally restricted by the super-hamiltonian and super-momentum
 constraints. In the classical dynamics the well-known Birkhoff theorem
 holds, which prohibits the existence of spherically symmetric
 gravitons. It is not certain whether even in quantum dynamics any
 interesting degrees of freedom for the exterior geometry of black
 holes can remain or not. However, we consider in this work, a spherically symmetric SdS black hole
 and require compatibility between the Hamiltonian and
 super-momentum constraints. We do not pursue the construction of
 mini-super-space models, which may be valid in the whole space-time.
 Our analysis is focused on quantization of the spherical hyper-surface $`t=Constant`,$ near the apparent horizon of a dynamical
 SdS black hole with a metric used in [24,25]. \\
  Details of the paper is as follows.
 In Sec.2 we derive WDW
equation of a SdS quantum black hole boundary metric near its
apparent horizon in time-independent regime. Then we use Taylor
series expansion of the self interacting mass function potential
and show that the WDW solution can be described via the quantum
harmonic oscillator. We obtained quantized mass eigenvalues of the
SdS quantum black hole. In Sec.4 we present a summary and
concluding remarks of this work.
 \section{Schwarzschild de Sitter quantum black hole}
Let us we start with a  dynamical massless scalar field $\psi$
minimally coupled with the Einstein-Hilbert gravity action in 4D
curved space time which is given  in units $G=c=\hbar=1$ as
follows.
 \begin{equation} I=\frac{1}{16\pi}\int
 dx^4\sqrt{g}(R-2\Lambda+\zeta
 g^{\mu\nu}\partial_{\mu}\psi\partial_\nu\psi
 )\end{equation}
where $\zeta$ and $`\Lambda`$ are the coupling constant and the
well defined cosmological constant.  $`g`$ is absolute value of
determinant of the background metric $`g_{\mu\nu}`$ and $`R`$ is
\textit{Ricci} scalar. Varying the above action with respect to
the metric components $`g^{\mu\nu}`$ we can obtain metric field
equation as follows.
\begin{equation} G_{\mu\nu}+\Lambda g_{\mu\nu}=-8\pi T_{\mu\nu}[\psi]\end{equation}
in which
\begin{equation}T_{\mu\nu}[\psi]=\frac{\zeta}{16\pi}\bigg[\partial_\mu\psi\partial_\nu\psi-\frac{g_{\mu\nu}}{2}
g^{\alpha\beta}\partial_{\alpha}\psi\partial_{\beta}\psi\bigg]\end{equation}
is matter stress-tensor. Setting $\Lambda=0$ and $\zeta=-16\pi$
the above equation has a static spherically symmetric
asymptotically flat exact solution called as Janis-Newman-Winicour
(JNW) solution as [26] \begin{equation}
ds^2=-\bigg(1-\frac{b}{r}\bigg)^sdt^2+\bigg(1-\frac{b}{r}\bigg)^{-s}dr^2+r^2\bigg(1-\frac{b}{r}\bigg)^{1-s}(d\theta^2+\sin^2\theta
d\varphi^2)\end{equation} with
\begin{equation}\psi=\frac{q}{b\sqrt{4\pi}}\ln\bigg(1-\frac{b}{r}\bigg)\end{equation} in which \begin{equation}s=\frac{2M}{b},~~b=2\sqrt{M^2+q^2}.
\end{equation} In the above metric solution $M$ and $q$ are called as the black hole mass and the charge of the scalar field respectively.
Setting $T_{\mu\nu}=0$ the
equation (2) has a spherically symmetric static metric solution as
follows [27] (see also [25]).
\begin{equation}
ds^2=-\bigg(1-\frac{2M}{r}-\frac{1}{3}\Lambda
r^2\bigg)dt^2+\frac{dr^2}{\bigg(1-\frac{2M}{r}-\frac{1}{3}\Lambda
r^2\bigg)}+r^2(d\theta^2+\sin^2\theta d\varphi^2).\end{equation}
The above metric  describes Schwarzschild static black hole in the
presence of the cosmological event horizon. The constants defined
by $M$ and $\Lambda$ are the black hole mass and the cosmological
parameters respectively. Defining a dimensionless cosmological
parameter
\begin{equation}\xi=\frac{\Lambda(2M)^2}{3}\end{equation} one can infer  that for $0<\xi<1$ the
above metric has two event horizon which are determined by solving
$1-\frac{2M}{r}-\frac{1}{3}\Lambda r^2=0.$ They are obtained as
$r_b\simeq2M(1+\xi)$ and $r_c\simeq\frac{2M}{\sqrt{\xi}}$ which
are named as the radius of the black hole and the cosmological
horizons respectively [7,8]. Size of the black hole horizon varies
between zero and a large scale of the cosmological horizon. If the
black hole horizon is much smaller than the cosmological horizon
so that $r_b\sim r_c$ for $\xi\sim0.46559,$ the effects of the
radiation coming from the cosmological horizon is negligible. In
this case it is named as a degenerate SdS black hole. Degenerate
solution in which the black hole has the maximum size is called
the Nariai solution [27]. In this solution the two horizons have
the same size and so the same temperature. Therefore it shall be
in thermal equilibrium when we choose $0\ll\xi<1.$ Intuitively one
would expect any slight perturbation of the geometry which makes
hotter the black hole with respect to its environment. Thus one
may suspect the thermal equilibrium of the Nariai solution (Large
scale black hole) to be unstable. We studied previously final
state of an evaporating quantum perturbed SdS black hole by
solving the backreaction equation in semiclassical quantum gravity
(the perturbative approximation) approach [7,8]. Mathematical
derivations of the backreaction equation in time-dependent and
time-independent regimes, predicts a minimal remnant stable static
SdS
 black hole final state. Instead of this perturbative approach we encourage to study stability of quantum evaporating SdS black hole
 in a non-perturbative canonical quantum gravity approach. To do so we should solve the WDW equation of the evaporating quantum SdS black hole
  as follows.  Because it is very difficult to study the
hamiltonian and supper-momentum constraints in the whole
space-time, here our analysis of the constraints is limited to the
region near the apparent horizon. This is useful to study the
black hole evaporation. Because we want to discuss how the
two-dimensional spatial area $4\pi \varphi^2$ of the apparent
horizon decreases. The apparent horizon location in the classical
regime is determined by the following null equation.
\begin{equation}
g^{\mu\nu}\partial_\mu\varphi\partial_\nu\varphi=0.\end{equation}
This is a natural spatial light-like surface where we can impose
the adequate boundary condition. In time independent regime we
assume that the evaporating SdS quantum black hole is formed as
\begin{equation}ds^2_{SdS}=-\bigg(1-\frac{2M(r)}{r}-\frac{\Lambda}{3}r^2\bigg)dt^2+\frac{dr^2}{\bigg(1-\frac{2M(r)}{r}-\frac{\Lambda}{3}r^2\bigg)}+r^2(d
\theta^2+\sin^2\theta d\varphi^2) \end{equation} in which mass of
the black hole $M(r)$ is changed vs $r$ because of the presence of
the quantum matter scalar field $\psi(r)$. Here we assume that the
quantum matter scalar field dose not changed the cosmological
parameter and so $\Lambda$ is still maintain as a constant.
Boundary counterpart of the above metric near the apparent horizon
$r=2M(r)$ reads
\begin{equation}\lim_{r\to2M(r)}ds^2_{SdS}\approx\frac{4}{3}\Lambda M(r)^2dt^2-\frac{dr^2}{\frac{4}{3}\Lambda M^2(r)}+4M^2(r)(d\theta^2+\sin^2\theta d\varphi^2).
\end{equation} The above metric corresponds to internal region of the evaporating SdS black hole near the horizon because its
signature is changed as $(+,-,+,+).$ This means that the time
coordinate $t$ behaves as spatial  coordinate while $r$ behaves as
the time coordinate. Similar to the mass function  the massless
scalar field should be defined versus the radial coordinate $r$
such that $\psi=\psi(r)$ in the time independent regime. Ricci
scalar for the boundary metric (11) reads
\begin{equation} R_{\mu}^{\mu}=R=-\frac{1}{2}\bigg(16\Lambda MM^{\prime\prime}+32\Lambda M^{\prime2}+\frac{1}{M^2}\bigg).\end{equation}
Substituting (11) and (12) and integrating on the two-sphere
region $\sin\theta d\theta d\varphi$ the action functional (1)
reads
\begin{equation} I=\int dtdt\mathcal{L}(M,M^{\prime}, \Lambda, \psi^{\prime})\end{equation}
where the lagrangian density $\mathcal{L}$ is obtained by
integrating by parts and eliminating divergence-less terms
$(M^3M^{\prime})^{\prime}$ as follows.
\begin{equation}\mathcal{L}(m,\dot{m},\dot{\sigma})=8m^2{\dot{m}}^2-\frac{3\zeta}{4}{\dot{\sigma}}^2-2m^2-\frac{1}{2}\end{equation}
where the dot $\dot{~}$ denotes to derivative with respect to
dimensionless radial coordinate $\rho=r\sqrt{\Lambda}$ and we
defined dimensionless mass function $m(\rho)$ and the field
$\sigma(\rho)$ as follows.
\begin{equation} m=\sqrt{\Lambda}M,~~~\sigma=\sqrt{\Lambda}\psi.\end{equation}
Applying the canonical momentum of the fields $\sigma$ and $m$ as
\begin{equation} \pi_{\sigma}=\frac{\partial\mathcal{L}}{\partial
\dot{\sigma}}=-\frac{3\zeta}{2}\dot{\sigma},~~~\pi_{m}=\frac{\partial\mathcal{L}}{\partial\dot{m}}=16m^2\dot{m}\end{equation}
and definition of the hamiltonian density
\begin{equation}\mathcal{H}=\pi_{\sigma}\dot{\sigma}+\pi_{m}\dot{m}-\mathcal{L}\end{equation} we obtain
\begin{equation} \mathcal{H}=\frac{\pi^2_{m}}{32m^2}-\frac{\pi_{\sigma}^2}{3\zeta}+2m^2+\frac{1}{2}.\end{equation}
To obtain the WDW wave equation of the boundary action functional
(11) near the black hole horizon we should substitute the Dirac`s
canonical quantization operators
\begin{equation} \hat{\pi}_{\sigma}=\frac{1}{i}\frac{d}{d\sigma},~~~\hat{\pi}_{m}=\frac{1}{i}\frac{d}{dm}\end{equation} into the super-hamiltonian
constraint (18) such that
\begin{equation}\bigg[\frac{d^2}{dm^2}-\frac{32m^2}{3\zeta}\frac{d^2}{d\sigma^2}-16m^2-64m^4\bigg]\Psi(m,\sigma)=0\end{equation}
where $\Psi(m,\sigma)$ is called as the WD probability wave
functional of the metric. It describes possible values of the
metric components (11) near the horizon when it takes on
particular values of the fields $\sigma$ and $m.$ We solve the WDW
wave equation (20) by applying the separation of variables method
as follows.
\begin{equation}\Psi(\sigma,\eta)=e^{i\sqrt{3\zeta(1+\zeta)}\sigma}R(m)\end{equation}
the WDW wave equation (20) reads
\begin{equation}\frac{d^2R}{dm^2}+V(m)R(m)=0\end{equation}
where we defined the self interaction mass potential $V(m)$ as
follows.
\begin{equation}V(m)=16(1+2\zeta)m^2-64m^4.\end{equation}
It is suitable to obtain Taylor series expansion of the above
potential about its minimum point as follows. Its minimum point is
obtained by solving the equation $\frac{dV(m)}{dm}=0$ as
\begin{equation} m^2_0=\frac{1+2\zeta}{16}\end{equation} for which one
can infer
\begin{equation}V(m_0)=192m_0^4,~~~\frac{d^2V}{dm^2}_{|_{m=m_0}}=-256m_0^2.\end{equation}
Up to higher order terms one can use (24) and (25) to obtain
Taylor series expansion of the potential (23) as follows.
\begin{equation} V(m)\approx 192m_0^4-126m_0^2(m-m_0)^2.\end{equation}
Substituting (26) and defining
\begin{equation} m-m_0=\kappa\chi,\end{equation} with \begin{equation}
\kappa^4=\frac{1}{126m_0^2},~~~192\kappa^2m_0^4=2n+1,~~~n=0,1,2,3,\cdots\end{equation}
the equation (22) reads
\begin{equation}\frac{d^2R(\chi)}{d\chi^2}+(1+2n-\chi^2)R(\chi)=0\end{equation}
which is similar to a quantum harmonic oscillator differential
equation. Its solutions is given versus the Hermit polynomials as
follows.
\begin{equation}R_n(\chi)=\frac{e^{-\chi^2/2}H_n(\chi)}{\sqrt{2^n\pi^2n!}}.\end{equation}
Its eigenvalues are obtained by solving (28) as follows.
\begin{equation}
m_{0}(n)=\bigg(\frac{2n+1}{12\sqrt{2}}\bigg)^{\frac{1}{3}}.
\end{equation}
Substituting (31) into the equations (15), (21) and (24) one can
obtain
\begin{equation} M(n)=\frac{1}{\sqrt{\Lambda}}\bigg(\frac{2n+1}{12\sqrt{2}}\bigg)^{\frac{1}{3}},\end{equation}
\begin{equation} \Psi_n(\chi,\sigma)=e^{if(n)\sigma}R_n(\chi)\end{equation} in which we defined
\begin{equation}
f(n)=\sqrt{\bigg(\frac{16(2n+1)}{\sqrt{2\sqrt{3}}}\bigg)^\frac{4}{3}-\frac{1}{4}}\end{equation}and
\begin{equation} \zeta(n)=8\bigg(\frac{2n+1}{12\sqrt{2}}\bigg)\frac{2}{3}-\frac{1}{2}.\end{equation}
We know that imaginary part of the WDW wave solution (33) dose not
have physical meaning while we can obtain maximal probability by
removing this imaginary part. This restrict us to choose some
quantized matter field as \begin{equation}
\psi(n,j)=\frac{\sigma(n,j)}{\sqrt{\Lambda}}=\frac{2j\pi}{f(n)},~~~j=0,1,2,3,\cdots.
\end{equation}Numerical values of the SdS quantum black hole
parameters $m_0(n),f(n)$ and $\zeta(n)$ are plotted in figure 1
versus the quantum numbers $n$. This process of the black hole
quantization may to be claim stability of an evaporating quantum
perturbed SdS black hole in high energy Planck regimes. It is in
accord to results of a perturbative approach of the problem which
previously is studied by the author [6,7].
\section{Concluding remarks}
Applying the canonical quantum gravity approach we solved the WDW
equation of the quantum SdS black hole metric near its apparent
horizon in presence of a massless quantum scalar matter field. WD
wave solution is obtained against the hermit polynomials and its
quantum numbers makes quantized mass of the quantum SdS black
hole. The most important message of this article is the
credibility of Penrose's cosmic censorship where causal
singularity of the SdS quantum evaporating black hole $r=0$ is
covered by the horizon. Its final state reaches to a minimal
remnant stable black hole where the positive cosmological constant
has important role for quantization of the SdS black hole.  As a
future work one can use our method given in the present paper to
quantize other forms of the black holes.

\begin{figure}\begin{center}
\hspace{0cm} \includegraphics[width=9cm]{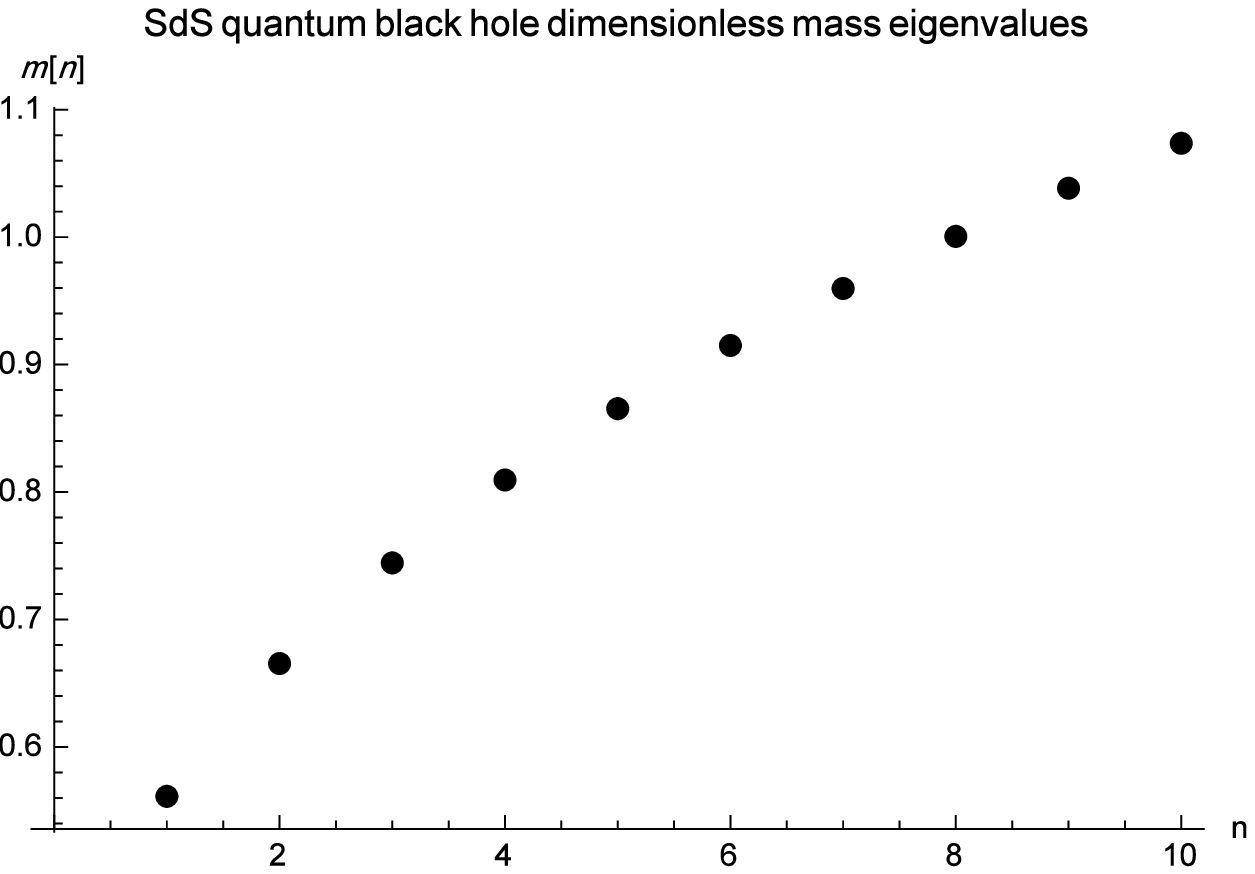}
\includegraphics[width=9cm]{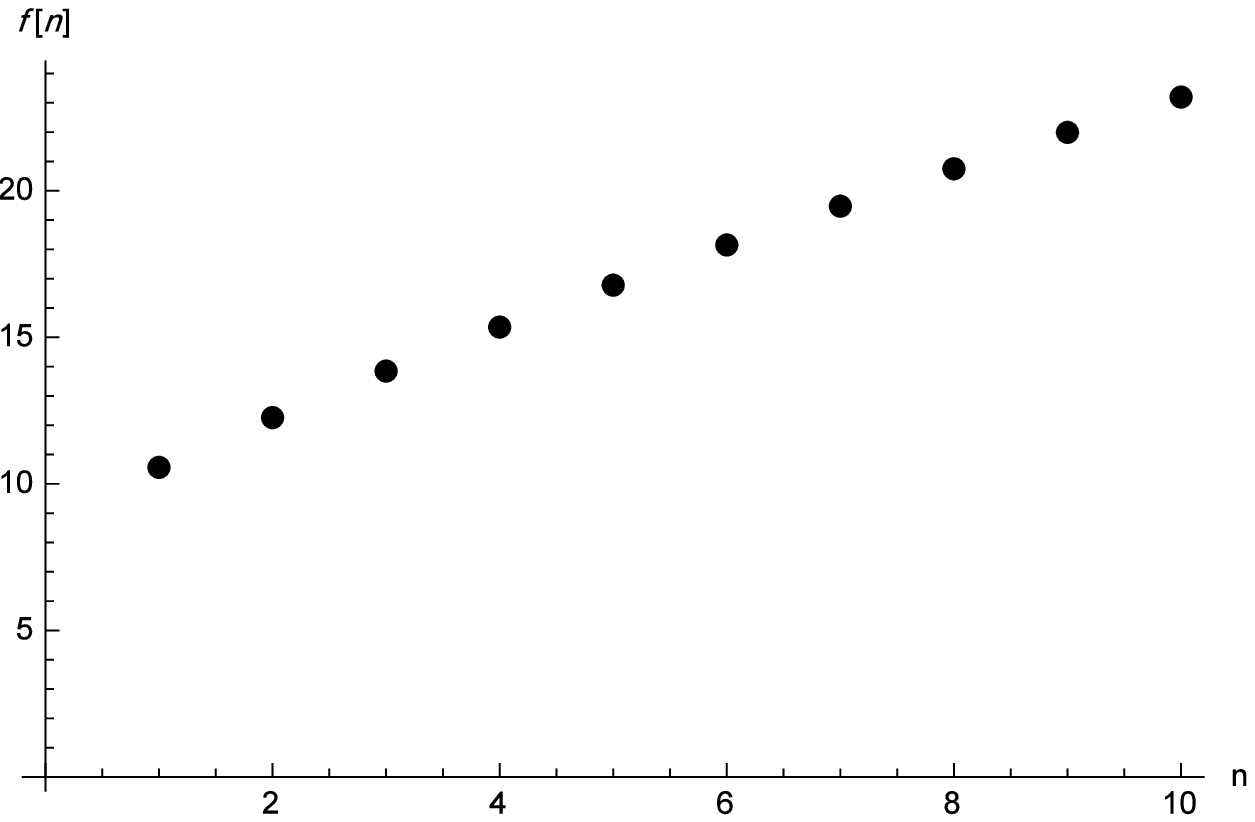}
\includegraphics[width=9cm]{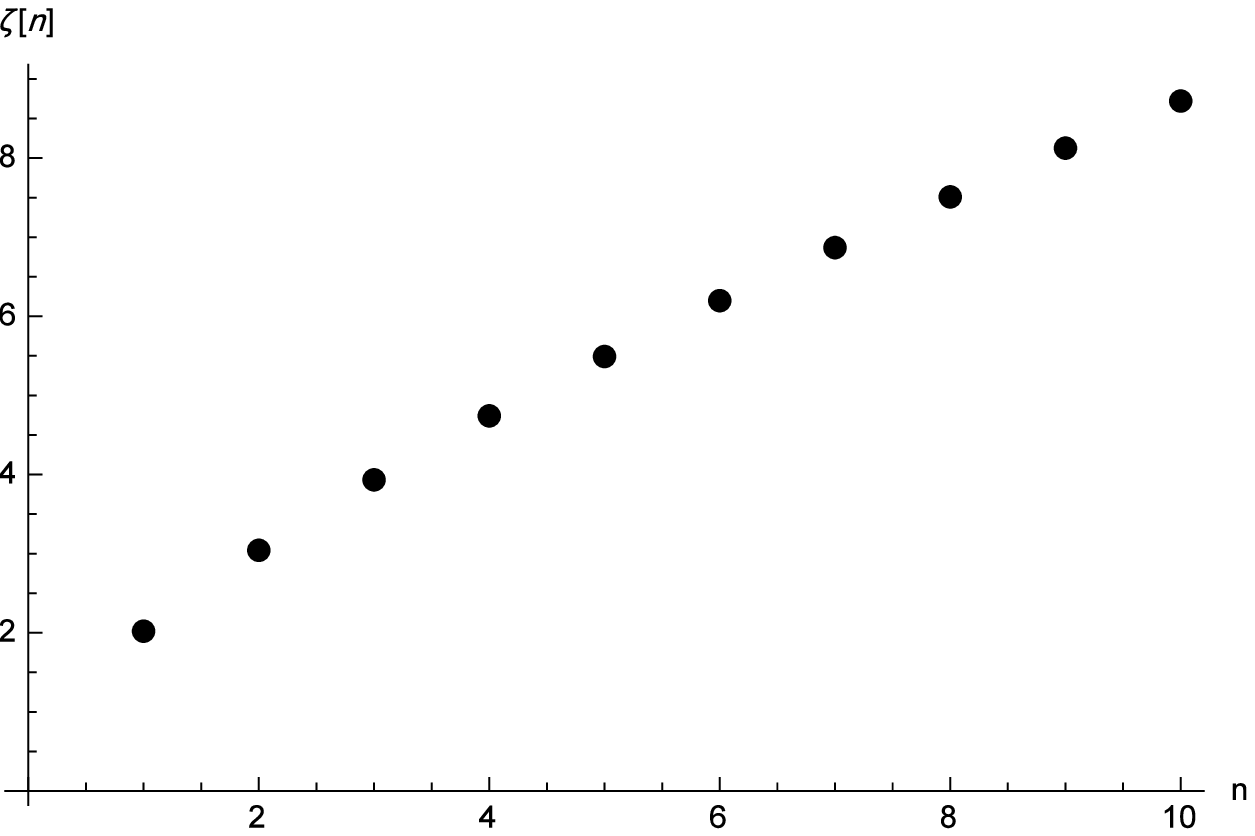}
\hspace{0cm} \caption{\label{mutbeta.figs.} Diagram of the SdS
quantum black hole parameters $m_0(n), f(n)$ and $\zeta(n)$ vs the
quantum numbers $n.$ }
\end{center}\end{figure}

\end{document}